\documentclass[aps,amsmath,twocolumn,amssymb,titlepage,10pt]{revtex4-1}
\usepackage[T1]{fontenc}
\usepackage[utf8]{inputenc}
\usepackage{amsmath}
\usepackage{braket}
\usepackage{amsfonts}
\usepackage{comment}
\usepackage{graphicx}
\usepackage{hyperref}
\usepackage[capitalize]{cleveref}
\usepackage{amssymb}
\usepackage{amsmath}
\usepackage{mathtools}

\begin{document}

\title{Phase transitions in the spin-1/2 Heisenberg antiferromagnet\\ on the dimerized diamond lattice}
\author{\firstname{Ronja} \surname{B\"{a}rwolf}
}
\affiliation{\mbox{Institute for Theoretical Solid State Physics, RWTH Aachen University, Otto-Blumenthal-Str. 26, 52074 Aachen, Germany}}
\author{\firstname{Alexander} \surname{Sushchyev}}
\affiliation{\mbox{Institute for Theoretical Solid State Physics, RWTH Aachen University, Otto-Blumenthal-Str. 26, 52074 Aachen, Germany}}
\author{\firstname{Francesco} \surname{Parisen Toldin}
}
\affiliation{\mbox{Institute for Theoretical Solid State Physics, RWTH Aachen University, Otto-Blumenthal-Str. 26, 52074 Aachen, Germany}}
\author{\firstname{Stefan} \surname{Wessel}
}
\affiliation{\mbox{Institute for Theoretical Solid State Physics, RWTH Aachen University, Otto-Blumenthal-Str. 26, 52074 Aachen, Germany}}

\begin{abstract}
Using a combination of unbiased quantum Monte Carlo simulations and a decoupled dimer mean-field theory, we investigate the thermal and quantum phase transitions of the spin-1/2 Heisenberg model on the dimerized diamond lattice. We find that at sufficiently strong dimerization the system exhibits a quantum disordered ground state, in contrast to the antiferromagnetic phase stabilized at weak dimerization. We determine the quantum critical point and examine the thermodynamic responses in both regimes.  The  ratio for the critical interdimer ($J_c$) to intradimer ($J_D$) coupling is obtained as $J_c/J_D = 0.3615(5)$. Our results show that the decoupled dimer mean-field theory well captures  the competition between the local singlet formation and the antiferromagnetic ordering tendency, and thus provides an appropriate  qualitative description of this three-dimensional quantum magnet, in contrast to the conventional mean-field decoupling. We also examine the differences in the ordering temperatures between the antiferromagnetic and the ferromagnetic spin-1/2 Heisenberg model on the  diamond lattice based on quantum Monte Carlo simulations. 
\end{abstract}
\maketitle

\section{Introduction}\label{Sec:Introduction}
Thermodynamic properties of three-dimensional quantum magnets are often well described by semi-classical or mean-field theory approaches. However, the underlying magnetic ordering tendencies can be strongly suppressed by frustrated interactions or enhanced quantum fluctuations resulting, e.g., from the proliferation of local singlets. In the past years, the latter scenario has indeed  been  investigated in various model systems -- both experimentally as well as using theoretical and computational approaches. A prominent  Cu-based spin-1/2 compound with suppressed magnetic order due to  a strong dimerization in the magnetic exchange interactions is ${\mathrm{T}\mathrm{l}\mathrm{C}\mathrm{u}{\mathrm{C}\mathrm{l}}_{3}}$, which furthermore exhibits a pressure-induced quantum phase transition towards a high-pressure antiferromagnetic state~\cite{Ruegg2004,Ruegg2005,Ruegg2008} -- indeed, pressure can be used to controll the relative dimerization strength in this material.
A faithful theoretical description of this compound requires to account for the exchange anisotropies that break the SU(2) symmetry of the most basic Heisenberg spin exchange interaction. The thermodynamic properties and the critical behavior in  coupled-dimer spin-1/2 systems have however also  been intensively examined in fully SU(2)-invariant models, such as dimerized Heisenberg antiferromagnets on the simple cubic and the bicubic lattice~\cite{Nohadani2005, Qin2015}. These investigations mainly focused on exploring universal properties at the thermal and quantum phase transitions, and elucidated the evolution of the magnetic excitations across the quantum phase transitions that separate quantum disordered and antiferromagnetic regimes~\cite{Lohoefer2017,Quin2017,Scammell2017}. 
In terms of the local connectivity, both the simple cubic lattice and the bicubic lattice, which is formed by two interwoven simple cubic lattices, are characterized by comparably large coordination numbers of $z=6$ and $7$, respectively. 

\begin{figure}
  \includegraphics[width=0.8\linewidth]{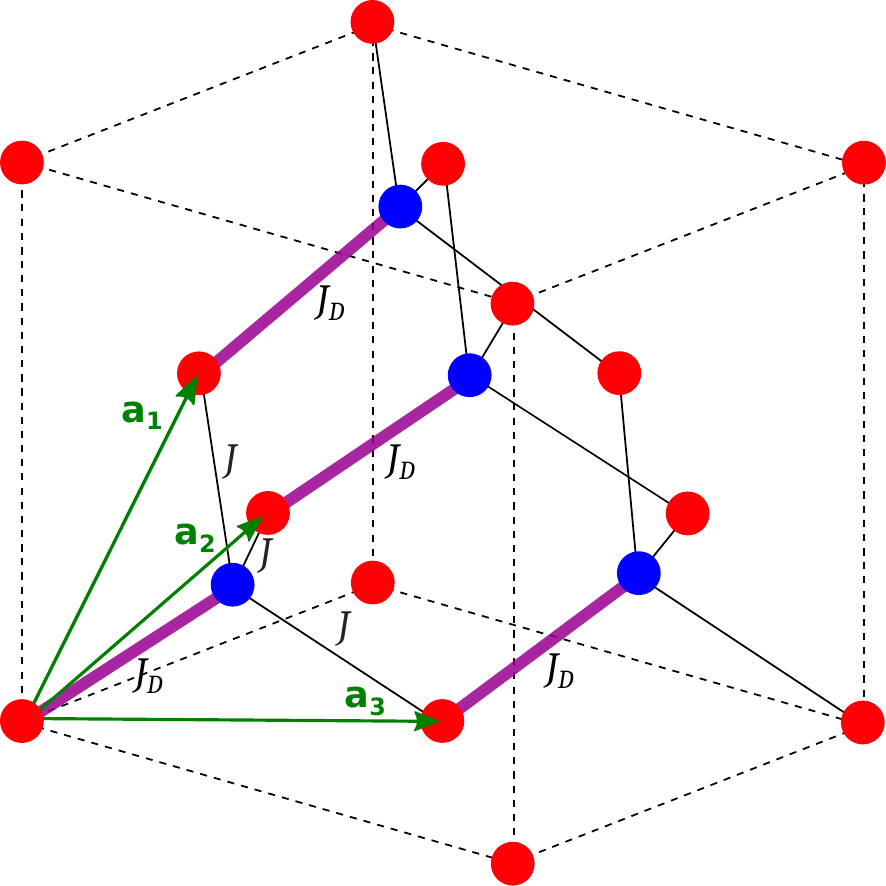}
  \caption{Illustration of the dimerized diamond lattice with lattice sites from the $A$ ($B$) sublattice shown in red (blue). Fat (thin) lines indicate  dimer (interdimer) bonds, with exchange couplings $J_D$ ($J$). Each lattice unit cell contains a single dimer bond. Also indicated are the  lattice vectors $\mathbf{a}_1$, $\mathbf{a}_2$, and $\mathbf{a}_3$, and the conventional unit cell by dashed lines.
  }
  \label{Fig:lattice}
\end{figure}

Here, we  will explore similar physics on the diamond lattice. It has a smaller local coordination of  $z=4$, which reduces magnetic ordering tendencies. However, due to its unfrustrated, bipartite nature, the antiferromagnetic N\'eel state can still be realized both in the ground state as well as at finite temperatures in this three-dimensional system (note that also the square lattice has $z=4$, but in two dimensions long-range order is destroyed by thermal fluctuations for the Heisenberg model).  Indeed, the N\'eel temperature of the  spin-1/2 Heisenberg antiferromagnet on the uniform diamond lattice (i.e., in the absence of any dimerization) as been estimated from high-temperature series expansion calculations as $T_N=0.531(1) J $, where $J$ denote the exchange interaction strength~\cite{Oitmaa2018}.
Moreover, as a face-centered cubic lattice with a two-site basis, the diamond lattice  can be  directly dimerized  upon increasing the bond strength within each unit cell, as illustrated in Fig.~\ref{Fig:lattice}. 

In the following, we therefore explore the magnetic properties of the antiferromagnetic spin-1/2 Heisenberg model on the dimerized diamond lattice. 
Based on unbiased quantum Monte Carlo (QMC) simulations using the stochastic series expansion approach with directed loop updates~\cite{Sandvik1999}, we refine the value of the N\'eel temperature for the uniform case and   locate the quantum phase transition that separates the antiferromagnetic regime  from the quantum disordered regime for sufficiently strong dimerization. In the latter phase, the system is  characterized by the dominant formation of local singlets on the strong dimer bonds. We also examine the thermodynamic properties in these different regimes. Our numerical analysis is complemented by a decoupled dimer-based mean-field theory (DDMFT) description of this system, which we detail in the following section, and which is similar in spirit to other dimer-based mean-field theory approaches~\cite{Tachiki1970,Tachiki1970b,Sachdev1990}.

Quite generally, the conventional, site-based mean-field theory of the Heisenberg model fails to account for the formation of the quantum disordered ground state that emerges upon increasing the dimerization strength. Instead, within the DDMFT theory, we treat only the interdimer spin exchange interactions within the mean-field approximation, thereby keeping intact the local quantum fluctuations within the dimers. Such a procedure allows us to account for both the antiferromagnetic ordering tendency as well as the local singlet formation within a comparably simple theoretical framework. In fact, the limit of vanishing interdimer coupling is trivially described exactly within this ansatz.  
As we verify below, within the DDMFT, the regime of strong dimerization and the nontrivial location of the quantum critical point are indeed captured accordingly, as is the thermodynamic response up to intermediate dimerization strength. 

The remainder of the article is organized as follows: in the following Sec.~\ref{Sec:DDMFT}, we introduce the considered quantum spin model and the DDMFT. Our results from QMC simulations on the quantum phase transition and the thermodynamic properties are presented in Sec.~\ref{Sec:QPT} and Sec.~\ref{Sec:TP}, respectively. A comparison between the QMC and DDMFT findings is provided in Sec.~\ref{Sec:Comparison}, and final conclusions are given  in Sec.~\ref{Sec:Conclusions}.

\section{Model and DDMFT}\label{Sec:DDMFT}
In the following, we consider the spin-1/2 Heisenberg model on the dimerized diamond lattice illustrated in Fig.~\ref{Fig:lattice} and described by the Hamiltonian
\begin{eqnarray}
H&=& J_D \sum_i \mathbf{S}_{i,A}\cdot \mathbf{S}_{i,B}\\
& & +\: J \sum_i  \mathbf{S}_{i,A} \cdot (\mathbf{S}_{i+\mathbf{a}_1,B}+\mathbf{S}_{i+\mathbf{a}_2,B}+\mathbf{S}_{i+\mathbf{a}_3,B})\:,\nonumber
\end{eqnarray}
where $\mathbf{S}_{i,A}$ ($\mathbf{S}_{i,B}$) denotes a spin-1/2 degree of freedom on sublattice $A$ ($B$) within the unit cell at position $\mathbf{r}_i$. Furthermore,  $i+\mathbf{a}_l$ denotes the unit cell at $\mathbf{r}_i+\mathbf{a}_l$, where $\mathbf{a}_l$, $l=1,2,3$ denote the three primitive lattice vectors (cf. Fig.~\ref{Fig:lattice}). The interaction within the dimers is denoted by $J_D$, and $J$ is the interdimer coupling. 
In the absence of dimerization, i.e., for $J=J_D$, the uniform diamond lattice is recovered. In the QMC simulations, we consider finite-size lattices with periodic boundary conditions, and the same linear extend $L$ is taken in each lattice direction, resulting in $N=2L^3$ lattice sites. We performed QMC simulations for systems sizes up to $L=40$. The results from these calculations will be discussed in the following sections. 

In addition to the QMC simulations, we consider here a cluster mean-field theory that is appropriate for bipartite dimerized lattices, the DDMFT. Related dimer-based mean-field theories have  previously been used to describe ground states of (frustrated)  quantum magnets~\cite{Sachdev1990} and coupled spin dimer systems in applied magnetic fields~\cite{Tachiki1970,Tachiki1970b}. 
In  conventional, site-based  mean-field theory of magnetic systems, all interaction terms $\mathbf{S}_j \cdot \mathbf{S}_k $ between two spins, denoted $\mathbf{S}_j$ and $\mathbf{S}_k$, are decoupled according to the mean-field approximation
as
$
\mathbf{S}_j \cdot \mathbf{S}_k \approx \mathbf{S}_j\cdot \langle \mathbf{S}_k  \rangle +
\mathbf{S}_k\cdot \langle \mathbf{S}_j  \rangle -
\langle \mathbf{S}_j  \rangle \cdot \langle \mathbf{S}_k  \rangle.
$
Within the DDMFT, we perform this decoupling  only for the interdimer bonds of strength $J$, thereby keeping for the intradimer bonds of strength $J_D$ the full quantum dynamics. Based on this approximation, the DDMFT thus leads to a description of the original quantum spin model in terms of an effective decoupled dimer system. Here, the two mean-fields $\mathbf{m}_A=\langle \mathbf{S}_{i,A}\rangle$ and $\mathbf{m}_B=\langle \mathbf{S}_{i,B}\rangle$ can be taken uniform, based on the fact that the antiferromagnetic state on the diamond lattice does not break the lattice translational symmetry. These two self-consistency equations can be readily solved by numerical iteration from the exact solution of the single dimer system, based on a full diagonalization of the effective dimer model Hamiltonian (corresponding to a $4 \times 4$ matrix in general). The latter is defined by the mean-field Hamiltonian
\begin{eqnarray}
H_\mathrm{DD}&=& \sum_i \left[\:  J_D  \: \mathbf{S}_{i,A}\cdot \mathbf{S}_{i,B}\right. \\
& & \left. +  \:3 J \: (\mathbf{S}_{i,A} \cdot \mathbf{m}_B + \mathbf{S}_{i,B} \cdot \mathbf{m}_A  - \mathbf{m}_A \cdot  \mathbf{m}_B ) \: \right]\:,\nonumber
\end{eqnarray}
which apparently decomposes into decoupled dimers, with $H_\mathrm{DD}=H$ in the limit $J=0$.
The internal energy is obtained as 
$E=\langle H_\mathrm{DD} \rangle /N$ for the selfconsistent solution and the specific heat follows from the derivative $C=dE/dT$ with respect to the temperature $T$. In the following, we fix the Boltzmann constant $k_B=1$.
We note that due to the SU(2) symmetry in the absence of an applied magnetic field, the mean fields can be chosen to align with, e.g., the spin $z$-direction,  i.e., $\mathbf{m}_A=(0,0,m^z_A)$, and $\mathbf{m}_B=(0,0,m^z_B)$. This reduces the above vector selfconsistency equations to two equations for $m^z_A$ and $m^z_B$, respectively, where $m^z_A=-m^z_B\neq 0$ holds true in  the antiferromagnetic regime.

The application of a longitudinal field $h_z$ that couples to the spin $z$-component of the quantum spins  (such that $H_\mathrm{DD}\rightarrow H_\mathrm{DD} - h_z\sum_i [S^z_{i,A}+S^z_{i,B}]$)  allows for the calculation of the longitudinal uniform magnetic susceptibility $\chi^\parallel=\lim_{h_z\rightarrow 0} (d m_z/d h_z)$, where $m_z=(m^z_A+m^z_B)/2$ (the factor $1/2$ provides normalization per site). In order to access the transverse susceptibility $\chi^\perp$, we instead apply a transverse field $h_x$ that couples to the spin $x$-component. For this calculation, we thus require in addition the spin $x$-components of the mean fields,  i.e., $\mathbf{m}_A=(m^x_A,0,m^z_A)$, and $\mathbf{m}_B=(m^x_B,0,m^z_B)$, and obtain $\chi^\perp=\lim_{h_x\rightarrow 0} (d m_x/d h_x)$, where $m_x=(m^x_A+m^x_B)/2$. In practice, we evaluate the above derivatives numerically after solving the selfconsistency equations. The rotationally averaged magnetic susceptibility, relevant for, e.g., powder samples or a direct comparison to zero-field QMC simulations, is obtained as $\chi=(\chi^\parallel + 2\chi^\perp)/3$.

The results obtained from the DDMFT are compared to unbiased QMC data in Sec.~\ref{Sec:Comparison}. Before performing this comparison though, we first present in the following section the QMC results for the quantum phase transition that separates the antiferromagnetic regime for $J\sim J_D$ from the quantum disordered phase at low values of the coupling ratio $J/J_D$.

\section{Quantum Phase Transition}\label{Sec:QPT}

\begin{figure}[t]
  \centering
  \includegraphics[width=\linewidth]{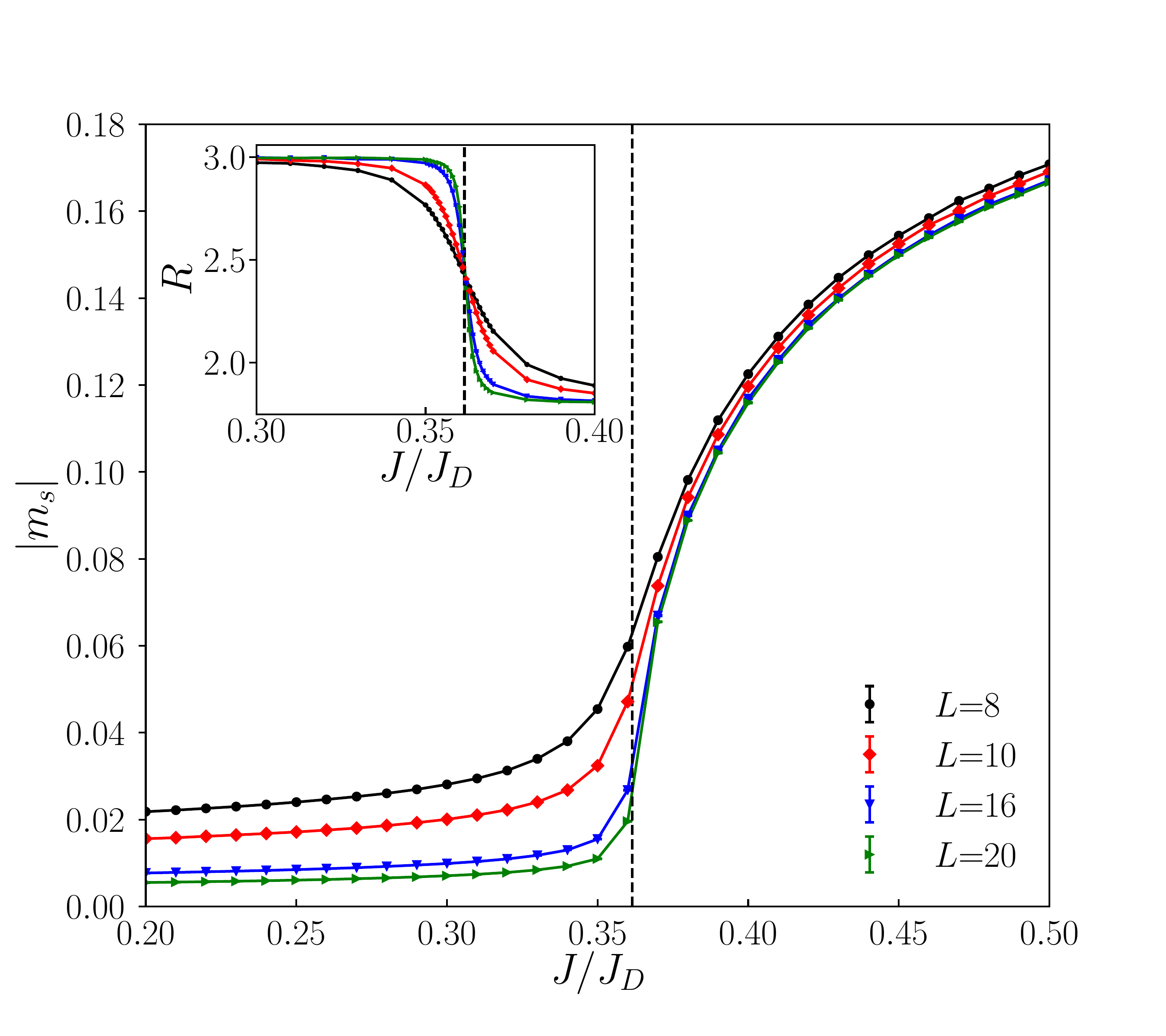}
  \caption{Dependence of the staggered magnetization $m_s$ on the coupling ratio $J/J_D$ for the antiferromagnetic spin-1/2 Heisenberg model on the dimerized diamond lattice for various  system sizes $L$. The inset shows the corresponding data for the Binder ratio $R$. Dashed lines in both panels indicate the location of the quantum critical point.
  }
  \label{Fig:QPT}
\end{figure}

In order to quantify the antiferromagnetic order that emerges at low temperatures in the region  $J\sim J_D$,  we calculate the corresponding antiferromagnetic structure factor 
\begin{equation}
    S_\mathrm{AF}=\frac{1}{N}\sum_{j,k} \epsilon_j  \epsilon_k \: \mathbf{S}_j\cdot \mathbf{S}_k,
\end{equation}
where $\epsilon_j=\pm 1$, depending on the sublattice ($A$ or $B$)  on  which the spin $\mathbf{S}_j$ is located, and  both  summations extend over all $N$ spins. Based on the structure factor, we estimate the staggered magnetization 
\begin{equation}
    m_s=\sqrt{S_\mathrm{AF}/N}.
\end{equation}
Furthermore, as detailed below, we can locate the quantum phase transition  out of the antiferromagnetic regime upon measuring the Binder ratio~\cite{Binder1981}
\begin{equation}
R=\frac{\langle M_s^4\rangle}{\langle M_s^2\rangle^2},
\end{equation}
where $M_s=\sum_j \epsilon_j S^z_j$ denotes the staggered moment.

In the main panel of Fig.~\ref{Fig:QPT}, the evolution of $m_s$ as a function of $J/J_D$ is shown for increasing system size. In these simulations, the temperature was scaled proportional to the inverse system size in order to probe ground state properties on these finite systems (more specifically, $T/J_D=(2L)^{-1})$.  This data already indicates that the antiferromagnetic order vanishes in the thermodynamic limit for sufficiently low values of $J\lesssim 0.36 J_D$. For a refined location of the quantum critical point, we identify the crossing point of the Binder ratio $R$ for different $L$, shown in the inset of Fig.~\ref{Fig:QPT}. Within the  resolution of the parameter scan, we can already estimate the critical interdimer coupling 
\begin{equation}
J_c = 0.3615(5) J_D
\end{equation} to a sufficiently high precision.
Based on previous studies of related bipartite three-dimensional coupled-dimer systems~\cite{Nohadani2005, Qin2015}, and as anticipated from the general quantum-to-classical mapping,  the critical point in such systems is well known to belong to the universality class of the classical $O(3)$ model in four dimensions, exhibiting logarithmic corrections atop the leading  mean-field scaling. The dynamical critical exponent $z=1$ was also already accounted for in the temperature scaling $T\propto 1/L$.

\section{Thermodynamic Properties}\label{Sec:TP}

\begin{figure}[t]
  \centering
  \includegraphics[width=\linewidth]{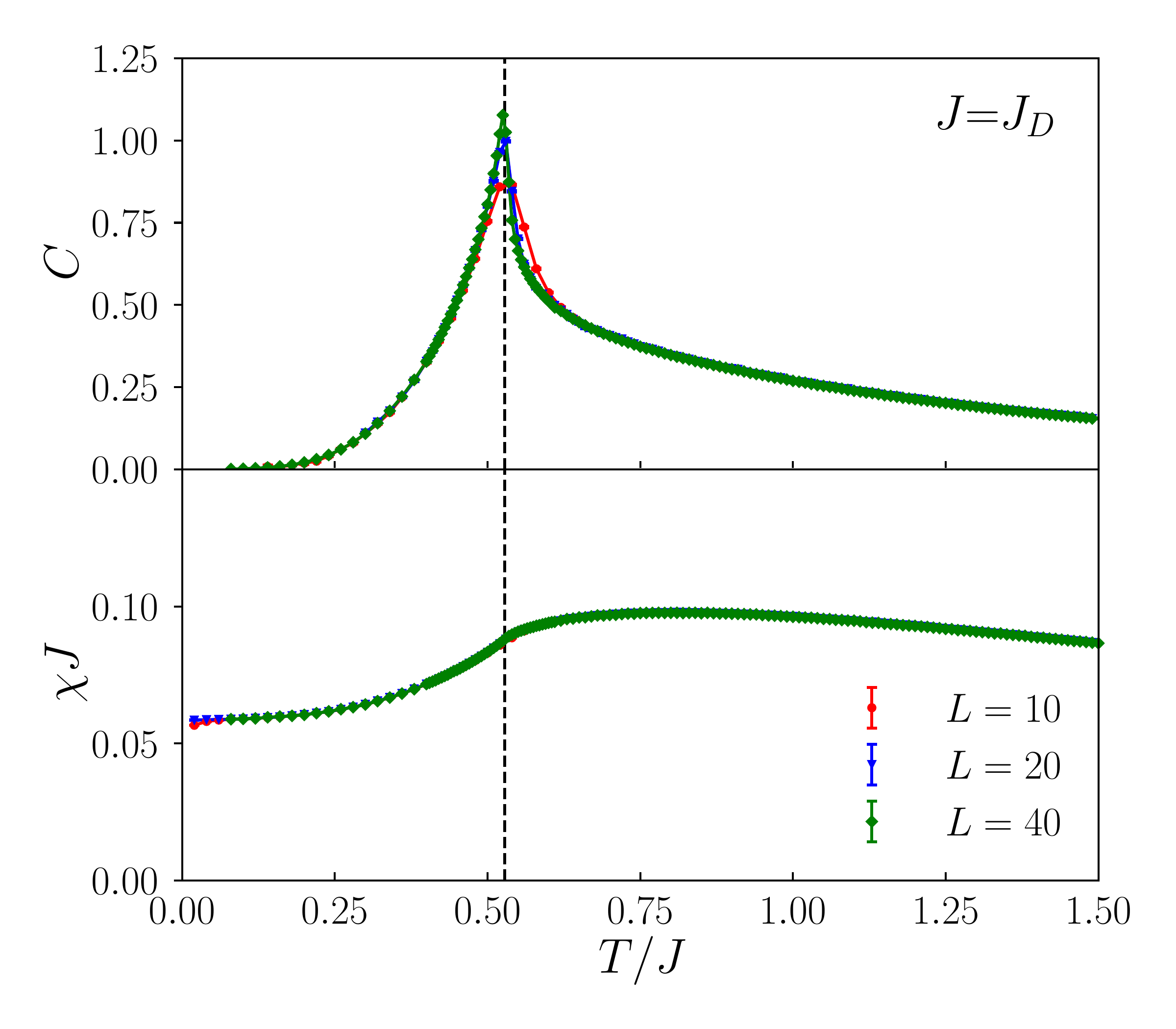}
  \caption{Temperature dependence of the specific heat $C$ and the susceptibility $\chi$ of the antiferromagnetic  spin-1/2 Heisenberg model on the diamond lattice at $J=J_D$  for various  system sizes $L$. Dashed lines in both panels indicate the location of the N\'eel temperature.
  }
  \label{Fig:Uniform}
\end{figure}

Before examining the dimerized diamond lattice, we consider the uniform case, i.e., $J=J_D$. For this system, an estimate for the N\'eel temperature
$T_N=0.531(1) J$ was reported based on high-temperature series expansion calculations~\cite{Oitmaa2018}. Here, we refine and complement this finding by performing QMC simulations in order to examine the thermal properties.

\begin{figure}[t]
    \centering
    \includegraphics[width=\linewidth]{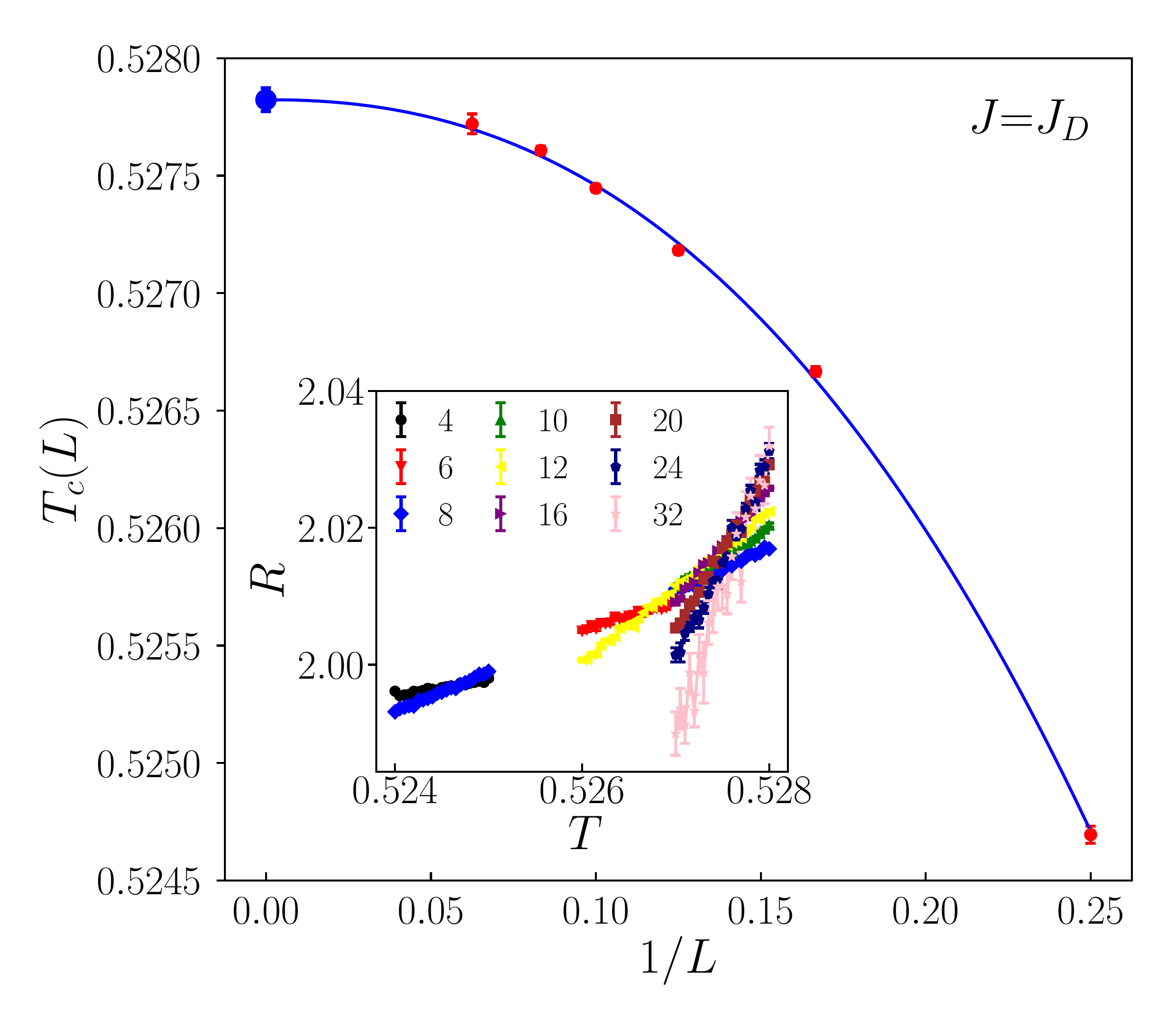}
    \caption{Finite-size extrapolation of the crossing points $T_c(L)$ of the Binder ratio $R$ for the antiferromagnetic spin-1/2 Heisenberg model on the diamond lattice ($J=J_D$). The inset shows the value of $R$ as a function of $J/J_D$ for various linear system sizes $L$, within the regime that is relevant for determining the crossing points of $R$ between linear system sizes $L$ and $2L$. 
    }
    \label{Fig:RUniform}
\end{figure}

Figure~\ref{Fig:Uniform} shows the temperature dependence of both the specific heat $C$ and the susceptibility $\chi$ for systems of different size $L$. A pronounced peak in the specific heat indicates the thermal ordering transition at the N\'eel temperature. For this case of $J=J_D$, 
we in addition  obtained a refined value of 
\begin{equation}
T_N= 0.52782(5) J
\end{equation}
from a further finite-size analysis of the Binder ratio $R$. More specifically, we monitor the temperature $T_c(L)$ of the crossing points  of $R$ for system sizes $L$ and $2L$, which is found to scale towards $T_N$ according to the standard finite-size scaling law $T_c(L) = T_N  + L^{-1/\nu}(c_1 L^{-\omega}+ c_2 L^{-2\omega})$~\cite{Barber83}, where $1/\nu = 1.406$ and $\omega = 0.78$ are the inverse of the correlation length and the leading correction exponents of the  three-dimensional $O(3)$ universality
class, respectively~\cite{Guida1998,Campostrini2002,Hasenbusch2001,Chester2021} ($c_1$ and $c_2$ are nonuniversal fit parameters). This is illustrated in Fig.~\ref{Fig:RUniform}. 

As seen from Fig.~\ref{Fig:Uniform}, the estimate of $T_N$ matches  well with the location of the specific heat peak. In the temperature dependence of the susceptibility $\chi$, shown in the same figure, we identify a broader maximum at about $T \approx  0.75 J$, as well as a slight kink at $T_N$. Moreover, the convergence of $\chi$ to a nonzero value upon further lowering $T$ is characteristic of the antiferromagnetic ground state (the slight drop of $\chi$ at small temperatures for the $L=10$ system is a residual finite-size effect).  

\begin{figure}
    \centering
    \includegraphics[width=\linewidth]{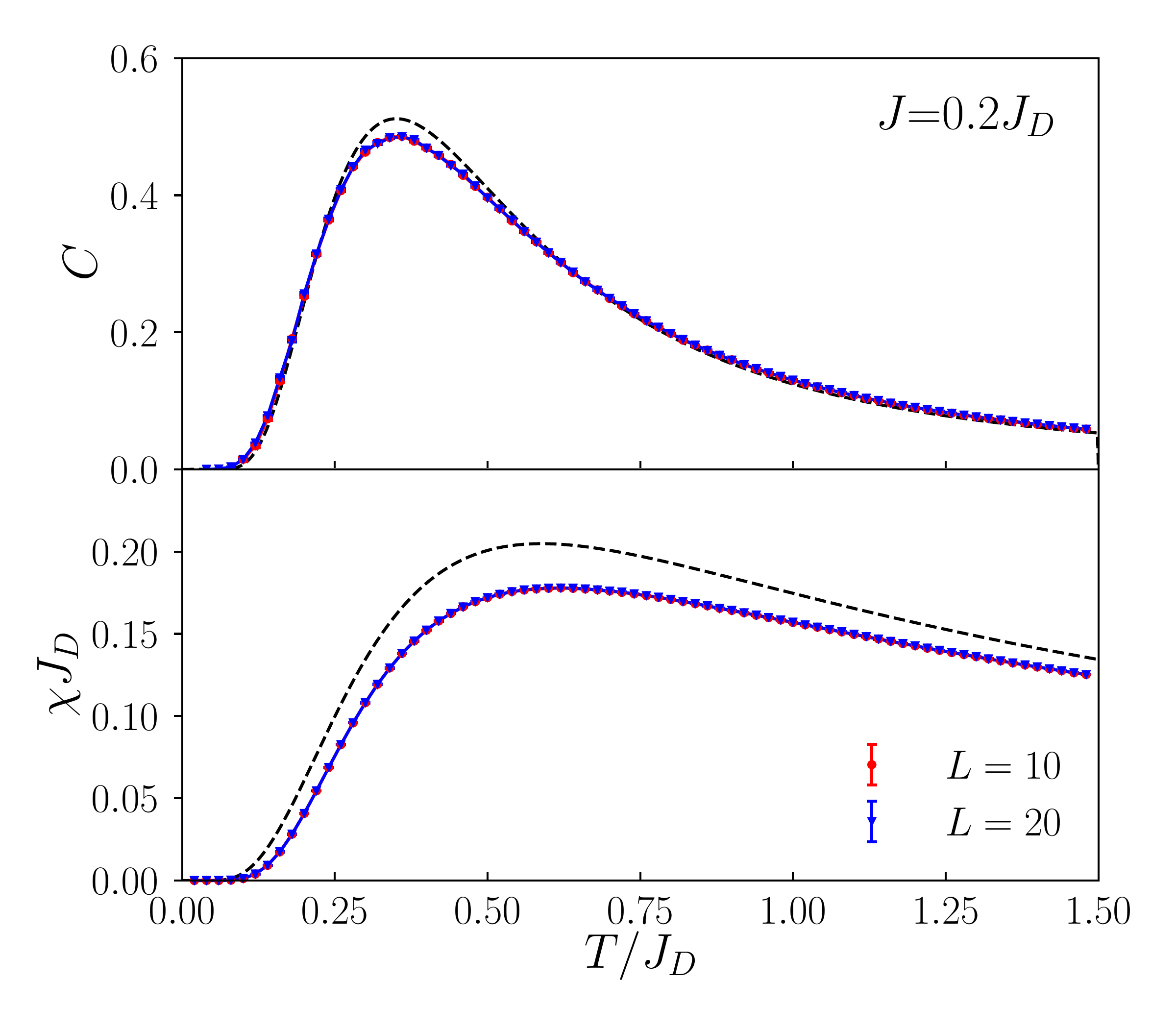}
    \caption{Temperature dependence of the specific heat $C$ and the susceptibility $\chi$ of the antiferromagnetic spin-1/2 Heisenberg model on the dimerized diamond lattice at $J=0.2 J_D$  for various  system sizes $L$. Also included by dashes lines are the results for isolated dimers, corresponding to $J=0$.}
    \label{Fig:J0.2}
\end{figure}

\begin{figure}
    \centering
    \includegraphics[width=\linewidth]{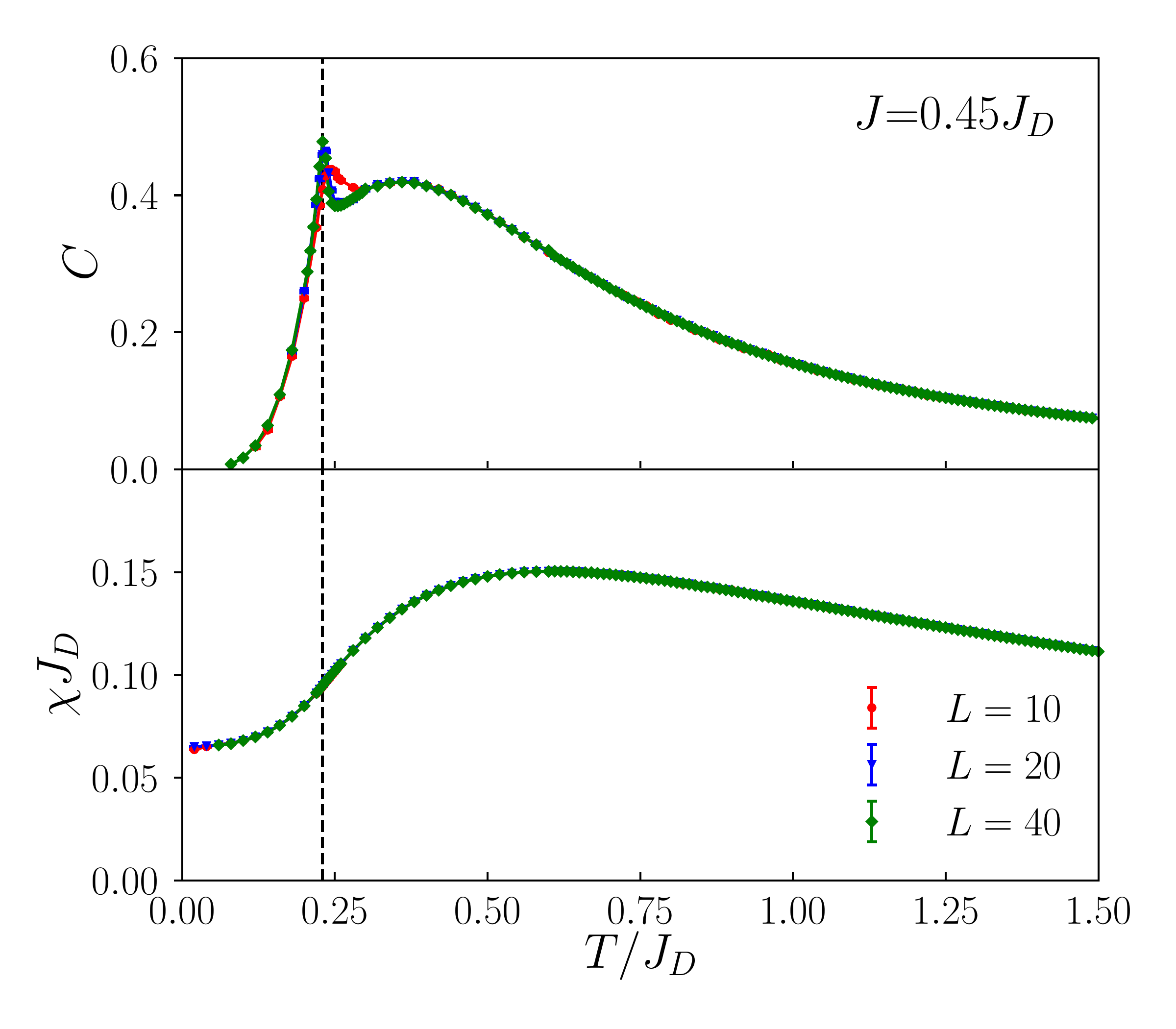}
    \caption{Temperature dependence of the specific heat $C$ and the susceptibility $\chi$ of the antiferromagnetic spin-1/2 Heisenberg model on the dimerized diamond lattice at $J=0.45 J_D$  for various  system sizes $L$. Dashed lines in both panels indicate the location of the N\'eel temperature.}
    \label{Fig:J0.45}
\end{figure}

\begin{figure}
    \centering
    \includegraphics[width=\linewidth]{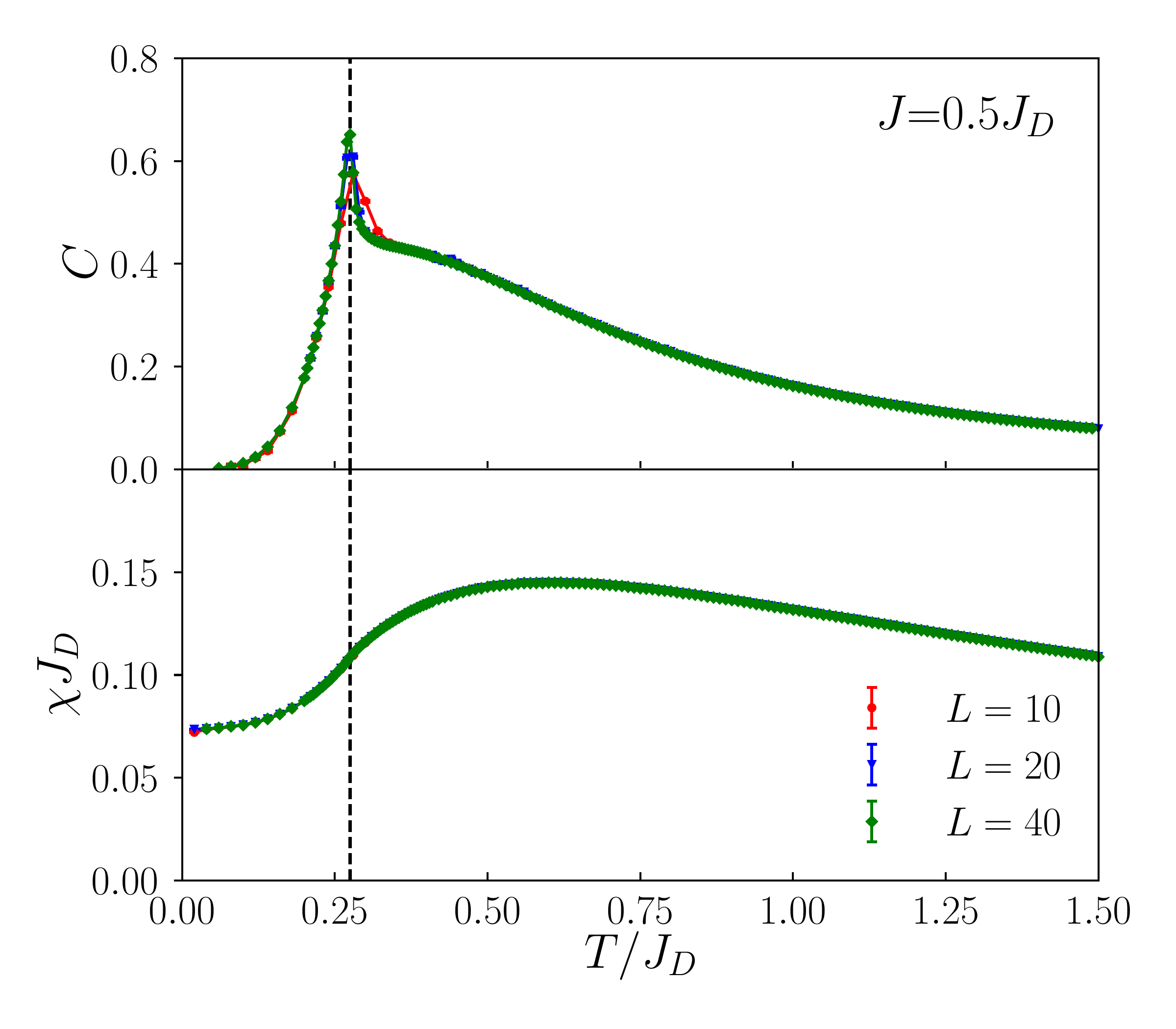}
    \caption{Temperature dependence of the specific heat $C$ and the susceptibility $\chi$ of the antiferromagnetic spin-1/2 Heisenberg model on the dimerized diamond lattice at $J=0.5 J_D$  for various  system sizes $L$. Dashed lines in both panels indicate the location of the N\'eel temperature.}
    \label{Fig:J0.5}
\end{figure}

We continue our investigation of the thermodynamic properties by now considering the case of the dimerized diamond lattice. More specifically, we show in Figs.~\ref{Fig:J0.2}, \ref{Fig:J0.45} and \ref{Fig:J0.5} the temperature dependence of both $C$ and $\chi$ for the cases of $J/J_D=0.2$, $0.45$ and $0.5$, respectively. 

For the case of $J =0.2 J_D$, the QMC data exhibits the characteristic features of the quantum disordered regime, in which the thermodynamic behavior deviates only weakly from the one of a system of completely decoupled spin dimers (corresponding to $J=0$), as can be seen by a comparison to the dashed lines in Figs.~\ref{Fig:J0.2}. Indeed, the value of $J/J_D=0.2$ lies below the quantum critical coupling ratio, and the activated thermodynamic response can be understood in terms of the singlet ground state and the gapped triplet excitations of  the spin dimers. The specific heat exhibits a broad maximum at $T\approx 0.35 J_D$, and the susceptibility at $T\approx 0.6 J_D$.  Both quantities exhibit essentially no finite-size effects due to the local character of the dimerized phase. 

Moving next to the case of $J=0.45 J_D$, i.e., beyond the quantum critical point, the specific heat now exhibits a two-peak structure, cf. Fig.~\ref{Fig:J0.45}.
The broad maximum at $T\approx 0.37 J_D$ is akin to the spin dimer physics, while the sharp  peak near $T\approx 0.23 J_D$ is due to the critical fluctuations near the thermal ordering transition at the  N\'eel temperature. The broad maximum in the susceptibility similarly reflects the response from the spin dimer system. 
In the low-$T$ regime, both $C$ and $\chi$ instead exhibit the characteristic behavior of the antiferromagnetically ordered regime, similar to the uniform system ($J=J_D$) that we discussed before.  

For $J=0.5 J_D$, cf. Fig.~\ref{Fig:J0.5}, the thermodynamic response is overall similar to the one at   $J=0.45 J_D$, however the N\'eel temperature has increased, such that the specific heat now exhibits only a single peat at $T\approx 0.27 J_D$, along with a broad shoulder at slightly elevated temperatures instead of forming a secondary maximum.
Further increasing $J$ then merely leads to a further  increase of the N\'eel temperature, eventually arriving at the previously discussed case of the uniform diamond lattice for $J=J_D$.

\section{Comparison to DDMFT }\label{Sec:Comparison}
Based on  QMC simulations, we identified both the thermal and the quantum phase transitions of the spin-1/2 Heisenberg model on the dimerized diamond lattice. As detailed in Sec.~\ref{Sec:DDMFT}, a simple approximate theory of such a bipartite coupled dimer system is obtained by performing a restricted mean-field decoupling of only the interdimer exchange interactions, thus keeping the full quantum description of the intradimer physics intact. In this section, we compare the results from these DDMFT calculations to the QMC results. 
Quantities obtained within the DDMFT will be denoted by a super- or subscript $\mathrm{DD}$, e.g., $C_\mathrm{DD}$ for the specific heat. 

\begin{figure}
    \centering
    \includegraphics[width=\linewidth]{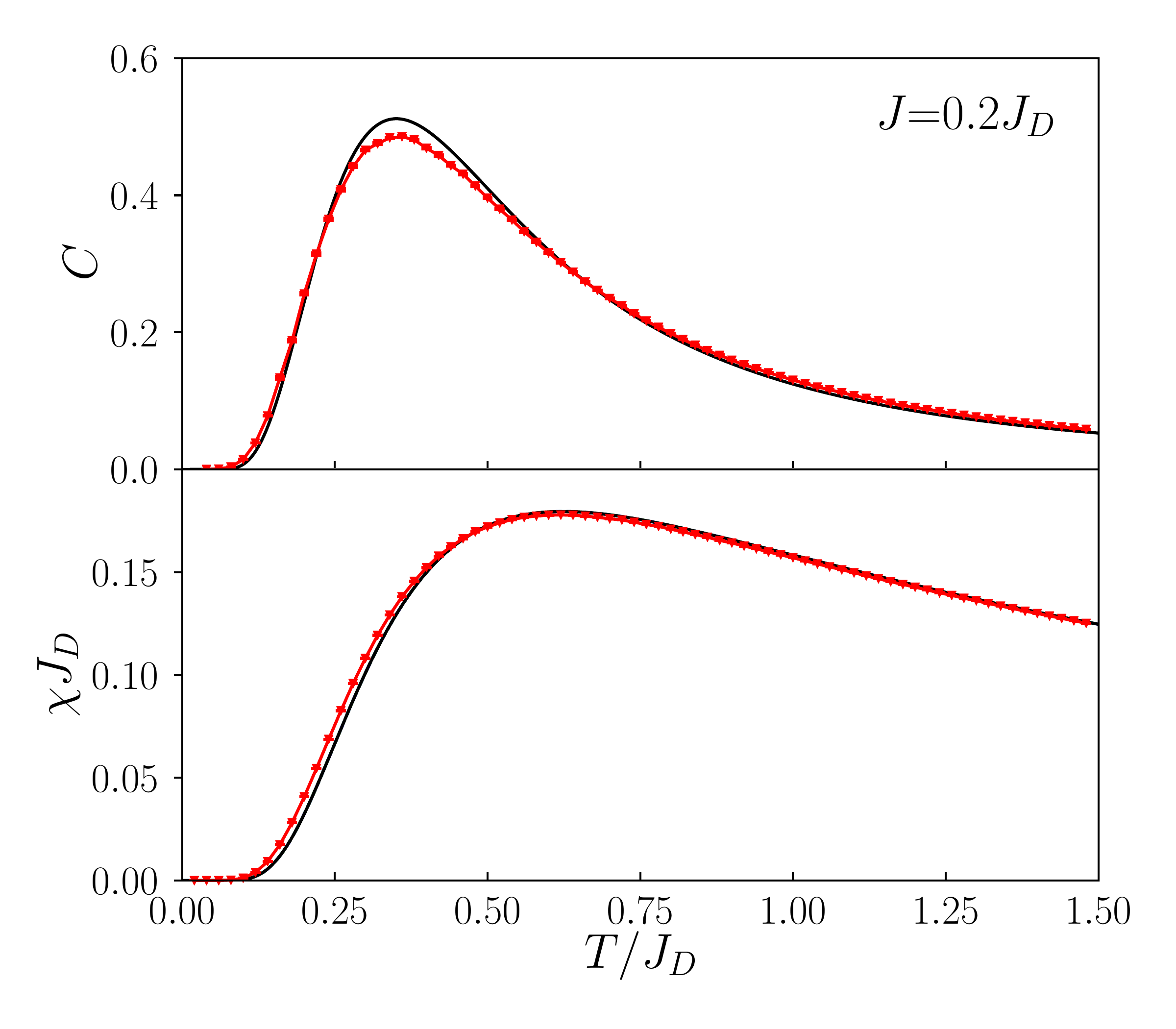}
    \caption{Comparison of the DDMFT results for the specific heat and the susceptibility  of the antiferromagnetic spin-1/2 Heisenberg model on the dimerized diamond lattice at $J=0.2 J_D$. Symbols denote QMC results and lines those from  DDMFT. 
   }
    \label{Fig:CompJ0.2}
\end{figure}

We first consider the case of $J=0.2J_D$, where the system resides in the quantum disordered phase according to the QMC results. In fact, also the DDMFT approach yields a nonmagnetic solution in this case, i.e., both mean-fields vanish. In Fig.~\ref{Fig:CompJ0.2}, we provide a comparison between the QMC and the DDMFT results, showing that overall the DDMFT captures the thermodynamic response rather well. Note that the specific heat $C_\mathrm{DD}$ obtained within the DDMFT is identical to the  specific heat in the decoupled dimer limit ($J=0$). Instead, the susceptibility $\chi_\mathrm{DD}$ obtained in DDMFT deviates from the decoupled dimer limit, because the magnetic field induces a finite polarization, which enters $H_\mathrm{DD}$ via the mean fields $\mathbf{m}_A$ and $\mathbf{m}_B$ (also note that $\chi^\parallel_\mathrm{DD}=\chi^\perp_\mathrm{DD}=\chi_\mathrm{DD}$ in the absence of antiferromagnetic order), yielding the usual random-phase-approximation (RPA)~\cite{Schulz1996} formula $\chi_\mathrm{RPA}=\chi_0/(1+3J\chi_0)$, where $\chi_0$ denotes the susceptibility of an isolated dimer ($J=0$), and the factor 3 in the denominator accounts for the  number of interdimer couplings of each spin (the formula results from combining $m^z_{A/B}=\chi_0 h^\mathrm{eff}_{z,A/B}$, where $h^\mathrm{eff}_{z,A/B}=h_z-3Jm^z_{B/A}$, with $m^z = \chi h_z$). 
In fact, the DDMFT susceptibility 
traces the QMC data rather closely. 

\begin{figure}
    \centering
    \includegraphics[width=\linewidth]{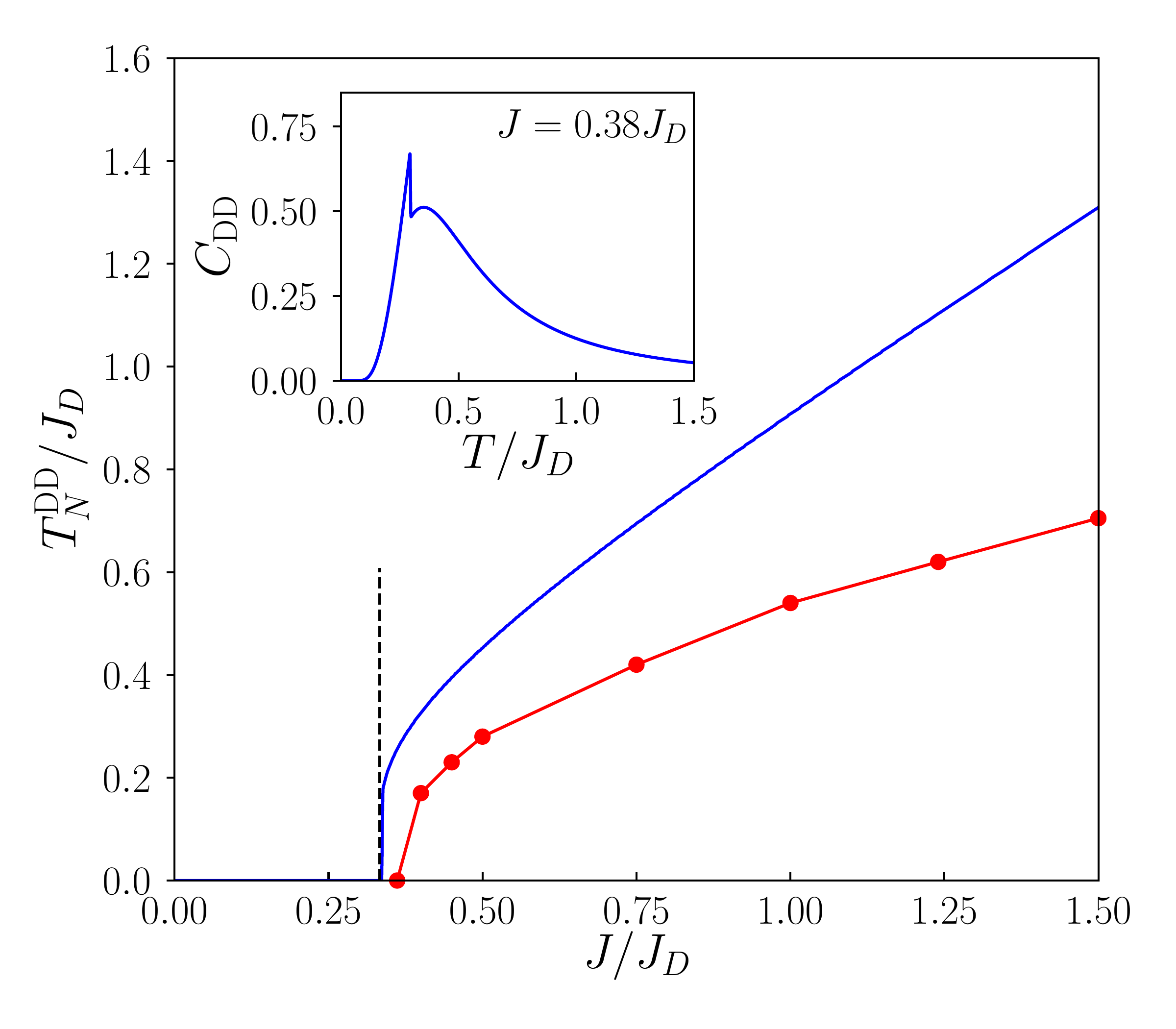}
    \caption{Dependence of the DDMFT values of the  N\'eel temperature $T_N^\mathrm{DD}$ for the antiferromagnetic spin-1/2 Heisenberg model on the dimerized diamond lattice on the coupling ratio $J/J_D$. Below the critical coupling of $J_c^\mathrm{DD}=J_D/3$, indicated by the dashed line, the ground state becomes quantum disordered.  For comparison, QMC results for the critical coupling $J_c$ and the N\'eel temperature $T_N$ are indicated by red circles, with error bars below the symbol size.  The inset shows the temperature dependence of the specific heat $C_\mathrm{DD}$ for the case of $J=0.38J_D$.  
    }
    \label{Fig:DDMFT}
\end{figure}

Upon monitoring the evolution of the DDMFT solution for increasing values of $J/J_D$, we identify the critical interdimer coupling strength of $J_c^\mathrm{DD}=J_D/3$, beyond which finite values of the mean-fields emerge at low $T$ (such finite values for the critical inter-dimer coupling are also obtained within the dimer-based bond-operator mean-field theory~\cite{Sachdev1990}). 
This is illustrated in the main panel of Fig.~\ref{Fig:DDMFT}, which shows the DDMFT values for the N\'eel temperature $T_N^\mathrm{DD}$ as a function of $J$. The position of $J_c^\mathrm{DD}$ is marked by a vertical line, below which $T_N^\mathrm{DD}$ vanishes. This value for the critical interdimer coupling follows also from locating the singularity in the RPA formula for the staggered susceptibility, $\chi^s_\mathrm{RPA}=\chi^s_0/(1-3J\chi^s_0)$, where $\chi^s_0$ denotes the staggered susceptibility of an isolated dimer ($J=0$), which approaches $J^{-1}_D$ in the low-$T$ regime, $\lim_{T\rightarrow 0} (\chi^s_0 J_D)=1$, from below. 

The DDMFT thus reproduces the quantum phase transition at a non-trivial value of the interdimer coupling (note that using a full mean-field decoupling of $H$, leading to conventional single-site mean-field theory, instead  locates the critical coupling at $J=0$, i.e., the system would be predicted to  order antiferromagnetically at low $T$ for any finite value of $J>0$). The DDMFT estimate of the critical coupling compares well to the QMC value, underestimating it by about 10\%. 
For $J$ beyond the critical coupling $J_c^\mathrm{DD}$, the specific heat $C_\mathrm{DD}$  exhibits a pronounced peak at the N\'eel temperature $T_N^\mathrm{DD}$. For $T>T_N^\mathrm{DD}$, $C_\mathrm{DD}$  is again identical to the specific heat of fully decoupled dimers, since the mean fields vanish in the paramagnetic regime. This results in a sudden drop of $C_\mathrm{DD}$ at $T_N^\mathrm{DD}$. For values of $J$ only slightly above the critical coupling $J_c^\mathrm{DD}$, such that the value of $T_N^\mathrm{DD}$ is smaller than the position of the maximum in the specific heat of the isolated spin dimer at $T/J_D=0.3515...$, the specific heat $C_\mathrm{DD}$ still exhibits a secondary maximum, namely the one related to  decoupled  dimers. This behavior is shown in the inset of Fig.~\ref{Fig:DDMFT} for the case of  $J=0.38 J_D$.

\begin{figure}[t]
    \centering
    \includegraphics[width=\linewidth]{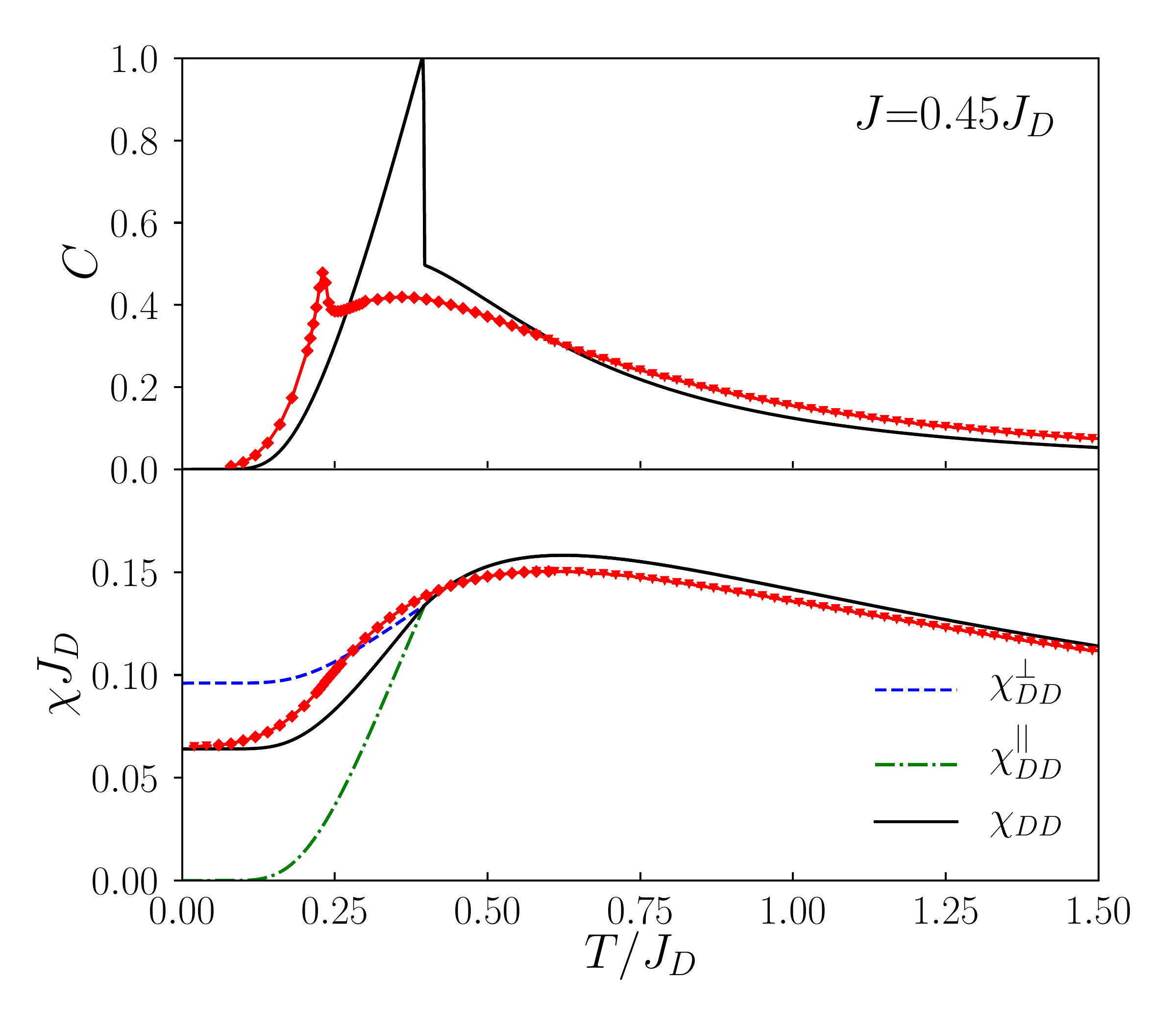}
    \caption{Comparison of the DDMFT results for the specific heat and the susceptibility  of the antiferromagnetic spin-1/2 Heisenberg model on the dimerized diamond lattice at $J=0.45 J_D$. Symbols denote QMC results and lines those from DDMFT. 
    }
    \label{Fig:CompJ0.45}
\end{figure}

As a further direct comparison of the DDMFT data to those from QMC, we consider in Fig.~\ref{Fig:CompJ0.45} the case of $J=0.45J_D$.
As expected, the DDMFT overestimates the N\'eel temperature and exhibits the aforementioned discontinuity in the specific heat at the single peak of $T_N^\mathrm{DD}$ for this value of $J$, while the QMC data for $C$ clearly indicates the presence of a secondary maximum. Also note that  the specific heat exponent $\alpha=0$ in mean-field theory, so that in DDMFT the specific heat does not exhibit a divergence. Similarly, 
since $\alpha=-0.1336(15)$~\cite{Campostrini2002} is negative for the three-dimensional O(3) universality class, there is also no divergence in $C$ at $T_N$, even in the thermodynamic limit. 
In order to compare the susceptibility data between the two approaches, one needs to keep in mind that the QMC calculations provide the rotationally averaged response $\chi$, which can thus be directly compared to $\chi_\mathrm{DD}$. Both quantities indeed exhibit very similar behavior, taking into account the difference in the  N\'eel temperature. While at low-$T$ the dominant contribution to the thermodynamic response is due to the Goldstone modes, the high-$T$ behavior is essentially that of the individual spins.  With respect to the two separate contributions $\chi^\parallel_\mathrm{DD}$ and $\chi^\perp_\mathrm{DD}$, the data in Fig.~\ref{Fig:CompJ0.45} exhibits the expected behavior, i.e.,  an activated (finite) low-$T$ response in the longitudinal (transverse) direction, while both functions merge in the paramagnetic regime above the N\'eel temperature.  

For even larger values of $J$, the DDMFT results still exhibit similar qualitative agreement to the QMC data, but quantitative deviations increase further, see Fig.~\ref{Fig:DDMFT}. For example, at $J=J_D$, DDMFT gives a critical temperature of $T_N^\mathrm{DD}\approx 0.903J$, compared to the actual value of  $T_N=0.52782(5) J$.  We finally mention that in the regime where $J$ becomes much larger than $J_D$, the DDMFT results approach towards those of the fully decoupled single-site mean-field theory, since the role of the differently treated $J_D$-terms diminishes. However, here we constrained ourselves  to the more interesting quantum regime of dominant dimer-coupling, and do not discuss further this large-$J$ regime.

\section{Conclusions}\label{Sec:Conclusions}
We used QMC simulations to identify the thermal and quantum phase transitions  of the spin-1/2 Heisenberg model on the dimerized diamond lattice. In particular, we identified the critical dimerization strength  -- in terms of $J_c/J_D=0.3615(5)$ --  below which a quantum disordered ground state of only weakly correlated spin dimers emerges. For the regime of dominant dimer couplings, we find that the DDMFT reflects the essential physics of this three-dimensional, bipartite system of coupled dimers.

\begin{figure}[t!]
    \centering
    \includegraphics[width=0.9\linewidth]{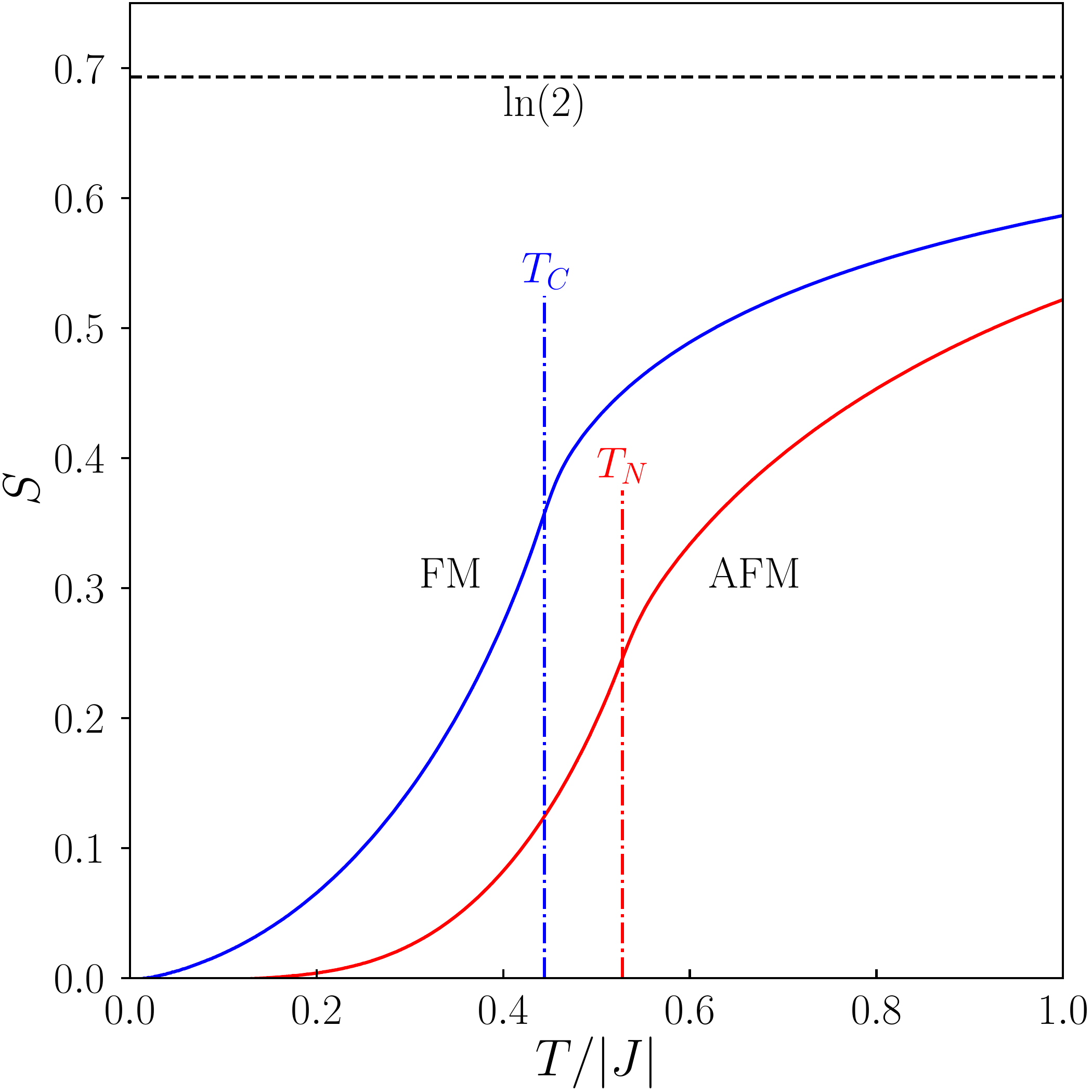}
    \caption{Temperature dependence of the entropy $S_\mathrm{AFM}$  ($S_\mathrm{FM}$) for the spin-1/2 Heisenberg antiferromagnet  (ferromagnet) on the diamond lattice ($J=J_D$).  Indicated by vertical  lines are the N\'eel and the Curie temperatures as well as the high-temperature limit $\ln(2)$ of the entropy (horizontal line).  
    }
    \label{Fig:S}
\end{figure}

For the case of the uniform diamond lattice, we provide a refined estimate for the N\'eel temperature, $T_N=0.52782(5) J$. As noted  in Ref.~\cite{Oitmaa2018}, this value of the critical (N\'eel) temperature $T_N$ is significantly larger than the corresponding (Curie) temperature  $T_C$ of the spin-1/2 Heisenberg ferromagnet on the same uniform diamond lattice. Indeed, Ref.~\cite{Oitmaa2018} reports a value of $T_C = 0.447(1) |J|$, where $J<0$ denotes the ferromagnetic spin interaction. Using the same QMC simulation and data analysis scheme from Sec.~\ref{Sec:TP} for the case of the ferromagnet (from considering the case of $J<0$), we obtain a similarly improved estimate for $T_C = 0.44447(5) |J|$, which confirms the observation and the value reported in Ref.~\cite{Oitmaa2018}.
In Tab.~\ref{Tab:Values} we collect together all high-precision critical point estimates from our QMC simulations. 
\begin{table}[t]
\caption{\label{Tab:Values} Critical points of spin-1/2 Heisenberg models on the diamond lattice, obtained from QMC simulations. }
\begin{ruledtabular}
\begin{tabular}{lll}
model & coupling sign & citical point location\\
\hline
uniform &  ferromagnetic & $T_C/|J| = 0.44447(5)$\\
uniform & antiferromagnetic & $T_N/J =0.52782(5)$\\
dimerized ($T=0$) & antiferromagnetic & $J_c/J_D= 0.3615(5)$\\
\end{tabular}
\end{ruledtabular}
\end{table}
The fact that indeed $T_N > T_C$ may at first sight appear surprising, given that one would consider the ferromagnetic, fully-aligned ground state more classical (and thus more robust with respect to thermal fluctuations) compared to the antiferromagnetic ground state, which  exhibits substantial quantum fluctuations.  In fact, this peculiar hierarchy of the critical temperatures is also observed for quantum Heisenberg models on  other bipartite three-dimensional lattices~\cite{Rushbrooke1963, Oitmaa2004, Wessel2010}.

As pointed out in Ref.~\cite{Wessel2010}, this observation can be rationalized in terms of the distinct dispersion relations between the ferromagnetic (quadratic) and antiferromagnetic (linear) case of the low-energy Goldstone modes in both systems. This distinction leads to a more pronounced proliferation of entropy upon increasing $T$ starting from the ferromagnetic ground state as compared to the antiferromagnetic case. In effect, this results in a larger critical entropy of the ferromagnet as compared to the antiferromagnet, i.e., $S_\mathrm{FM}(T_C)>S_\mathrm{AFM}(T_N)$ -- hence, the ferromagnetic state is indeed stable towards larger thermal fluctuations as quantified by the entropy. This behavior is illustrated for the specific case of the uniform diamond lattice in Fig.~\ref{Fig:S} (the entropy  was obtained from the QMC data of the energy for the $L=16$ system upon performing standard thermodynamic integration), and we expect it to apply rather generically to bipartite three-dimensional lattices, due to its symmetry-based character. Indeed, it is in  accord with previous observations~\cite{Rushbrooke1963, Oitmaa2004, Wessel2010}, and explains the generic finding  $T_N > T_C$ naturally in terms of the low-energy excitations of the magnetically ordered regime.

 Finally, we would like to mention that in recent years several compounds have been investigated that realize quantum spin systems with a diamond lattice geometry, of which some also form non-magnetic low-temperature states~\cite{Ge2017,Chamorro2018,Das2019,Abdeldaim2020,Kelly2022,Zager2022}.
 Here, we have examined a most basic scenario for such behavior, driven by the spin singlet physics due to a sufficiently strong dimerization in the exchange couplings. An alternative route towards non-magnetic ground states is provided by frustrated interactions, e.g., in the case where also exchange couplings beyond nearest-neighbor spins become relevant. Such extended couplings are apparently relevant for the investigated compounds as well. The QMC treatment of such frustrated quantum magnets is however well-known to be plagued in most cases by a serious sign-problem~\cite{Henelius2000,Troyer2005,Hangleiter2020,Hen2021}.  
While for  certain so-called fully frustrated models the sign problem can indeed be completely eliminated, cf.\ Refs.~\cite{Nakamura1998,Alet16,Honecker16,Weber2022,NgYang17,Stapmanns2018,Fan2024}, in general alternative approaches are required to quantitatively explore such systems theoretically. One possibility is the use of functional renormalization group methods for spin systems, which provide useful approximate results for several three-dimensional quantum magnets~\cite{Iqbal2016, Mueller2024}.  Certainly, the results reported here will be of value also for further probing and improving such calculation schemes toward  accurately treating quantum spin systems with  enhanced quantum characteristics.

\begin{acknowledgments}
We thank Y. Iqbal and J. Reuther for discussions and for the  motivation to examine the Heisenberg model on the  diamond lattice for different signs of the exchange constant, and A. Honecker for discussions  and for suggesting a dimer-based mean-field decoupling. Furthermore, we acknowledge computing time granted by the IT Center of RWTH Aachen University. F.P.T. is funded by the Deutsche Forschungsgemeinschaft (DFG, German Research Foundation), Project No. 414456783.
\end{acknowledgments}
\bibliography{main.bbl}

\end{document}